**Longitudinal Computer-Generated Holograms for Digital Frequency Control in Electronically Tunable Terahertz Lasers**

Subhasish Chakraborty[1], Owen P. Marshall[1], Md. Khairuzzaman[1], Harvey E. Beere[2] & David A. Ritchie[2]

[1]School of Electrical and Electronic Engineering, University of Manchester, Manchester M13 9PL, UK.

[2]Cavendish Laboratory, Department of Physics, University of Cambridge, Cambridge CB3 0HE, UK. Email: s.chakraborty@manchester.ac.uk

**A *transverse* computer-generated hologram (CGH) diffracts and provides flexible control of incident light by steering it to any point in the projected image plane – i.e. CGHs are able to direct the light to where it is needed and away from where it is not[1]. In addition, the number of resolvable points in the image projection plane is a function of the CGH's pixel count[2]. Here we report a *longitudinal* CGH (LCGH), a photonic structure, which swaps the ability to steer light toward fixed spatial points for digital control in the frequency domain. This is of particular interest in the context of tunable lasers. In this regard, an LCGH offers two important degrees-of-freedom (DOFs): 1) provides high-resolution *wavevector* or *k*-space resolution within the Brillouin zone; 2) enables full control to define or modify the reflectivity at each resolvable *k*-point, so attaining a target spectral response. We demonstrate the flexibility of our LCGH approach by achieving purely electronic tuning between six digitally-selected operating frequencies in a single-section terahertz (THz) quantum cascade laser (QCL)[3]. These switchable single-frequency devices will simplify combining the power and flexibility of**

**THz QCLs with spectroscopic applications, such as remote sensing, spectral analysis, and both security and medical imaging[4,5].**

Significant electronic tuning and single-frequency emission have conventionally been difficult to realise simultaneously in a single THz QCL[6-11]. Achieving multiple Bragg resonances within the gain bandwidth of a QCL – a prerequisite for switchable multi-colour operation[12] – presents particular technical difficulties at THz frequencies. For example, the multiperiodic grating structures widely used in lower-wavelength tunable semiconductor lasers, such as sampled gratings[12,13] and concatenated gratings with an aperiodic basis[14], would require total device lengths of centimetres if scaled to THz QCLs; a size difficult to fabricate and electrically power. Similarly, tunable coupled-cavity laser designs[15] are difficult to implement in surface plasmon (SP) waveguides[3] and to date have not been reported for THz QCLs. On the other hand, quasi-periodic gratings based on fixed mathematical sequences do not provide the required spectral flexibility[16], whereas an array of distributed feedback (DFB) lasers[17] would require additional driving electronics and collection optics. Electronically-switchable multi-colour THz QCLs therefore demand a fresh approach. The first DOF of an LCGH provides a solution. Here, a binary LCGH (Fig. 1a) is used to digitally encode the inverse Fourier transform (FT) of a desired spectral response function – in this case multiple, finely spaced, Bragg resonances within the gain bandwidth of a THz QCL (Figs. 1b and c). Digital frequency selection within an LCGH-QCL relies upon a one-to-one mapping between Bragg resonances and lasing modes (Figs. 1c and e), with current induced changes in the spectral gain (Fig. 1d) enabling switching between these modes, i.e. electronically-controlled discrete tuning.

In order to understand the first DOF, consider an LCGH of $2N$ pixels in the form of a spatial relative permittivity distribution $\varepsilon(z)$, where $L = N\Lambda$ is the total length and $\Lambda$ the minimum hologram-element separation. There exists an approximate FT relationship between $\varepsilon(z)$ and the spectral reflectivity response $\rho(k)$[18-20]:

$$\rho(k) \approx \tanh\left|\frac{1}{4n_{\text{eff}}^2}\int\limits_{-\infty}^{\infty}\left[\frac{\partial\varepsilon(z)}{\partial z}\right]e^{j2kn_{\text{eff}}z}\,dz\right| \tag{1}$$

Due to the pixelated nature of the spatial domain $z$, the wavevector $k$ is unique only over the interval $(0,k_{\text{B}})$, with a maximum $N$ number of resolvable $k$-points resulting in a density of states $\Delta k = (k_{\text{B}}/N)(n_{\text{eff}}/n_{\text{g}})$[21, 22], where $k_{\text{B}} = \pi/n_{\text{eff}}\Lambda$ is the wavevector corresponding to the edge of Brillouin zone, $n_{\text{eff}}$ is the effective modal refractive index, $n_{\text{g}}$ is the group refractive index and $c$ is the speed of light in vacuum. Note that $k_{\text{B}}$ ($f_{\text{B}} = ck_{\text{B}}/2\pi$) is also equivalent to the Bragg wavevector (Bragg frequency) of a uniform grating with a periodicity $\Lambda$. It is this FT basis that allows multiple Bragg resonances to be created with a minimum separation given by $\Delta k/k_{\text{B}} = 1/N$, neglecting dispersion.

Moving from a single strong resonance at $k_{\text{B}}$ (e.g. see the inset of Fig. 1, which was obtained using equation 1 for a uniform grating) to a multi-resonance response tends to reduce individual resonance strengths when $N$ and the refractive index perturbation $\Delta n$ remain unchanged ($|\Delta n| = 0.1$ and $N = 200$ in Fig. 1). Fortunately, this trade-off proves advantageous in achieving tunable LCGH-QCLs. During the course of this work it was discovered that weak individual resonances are a requirement for optimum tunability: the individual resonances must not be too strong that frequency migration is hindered under varying laser-driving conditions. When optimizing the LCGH spectral response (Figs. 1b and c), our goal was to limit Bragg resonance strengths to a level comparable to the waveguide facet reflectivity. To this end, the second DOF of an LCGH allows us to "throw-away" the remaining undesired strength from individual resonances to unused areas of the Brillouin zone. This is a powerful additional means to control resonance strengths as compared with changing $\Delta n$, or simply $N$. The LCGH capabilities, as highlighted by the two fundamental

DOFs outlined above, are what make this work intrinsically distinct from periodic[23], quasi-periodic[16], chirped[24], coupled-cavity[15] and superstructure gratings[12,13], or from concatenated gratings with an aperiodic basis[14].

All QCLs were fabricated from a molecular beam epitaxially grown GaAs/Al$_{0.15}$Ga$_{0.85}$As wafer, V557, with an 11.4 μm-thick active region based on reference 25. V557 was processed (Fig. 2 caption) into SP waveguides and cleaved into ~6 mm-long Fabry-Perot (FP) cavities. All devices displayed similar performance characteristics − as a typical example Fig. 2a shows the FP spectra of device $A$, recorded at four driving current densities. As expected, multiple longitudinal FP modes are present under all operating conditions, indicating an $n_g$ of 3.82. Device $A$ lased between 2.79 and 3.06 THz as the driving current density was increased from a threshold, $J_{th}$ of 106 A/cm$^2$ to the cessation of lasing at 265 A/cm$^2$. The FP spectra reveal the underlying blue-shift of the V557 gain-shape as a function of QCL driving current. Subsequently the target spectral response $\rho_{target}$ given in Figs. 1b and c was conceived in Fourier-space. It consists of three regions: (i) a region accommodating the underlying blue-shift of the gain-shape with the Bragg resonance strengths, so as to facilitate optimum tunability; (ii) a "throw-away" region (not necessarily contiguous with the first region), where an additional set of Bragg resonances are introduced; and (iii) a region which generally has no resonances (see Fig. 1b caption). Finally, to ensure highly delineated resonances every other resolvable $k$-point was chosen, resulting in a normalised target tuning step (TS$_{target}$) of $2 \times 1/N = 0.01$, using $N = 200$.

In order to generate the real-space lattice structure (i.e. the LCGH) satisfying $\rho_{target}$, we exploit equation 1. The pixelated nature of the real-space allows this design to be implemented using a discrete FT, specifically a fast Fourier transform (FFT). Identifying an "optimised" LCGH architecture is computationally non-trivial, particularly when $N$ is large. An FFT-based simulated annealing (SA) inverse optimisation algorithm was chosen, details of which, including the number of optimisation parameters, are described in references 26 and 27. The choice of algorithm is not critical - alternatives such as genetic algorithms[28] are equally applicable. We stress that the novelty of our work does not lie in the already well-established inverse design or Fourier-space engineering methodologies, but in the $\rho_{target}$ made possible by the LCGH DOFs.

An FEI Nova Nanolab 600 Focussed Ion Beam (FIB) system (30 keV Ga ions, 50 nm nominal spot diameter, 1 nA beam current) was used to transfer each hologram-element (Fig. 1a) as a discrete reflection site into the QCL ridge by milling 100 $\mu$m wide slits, penetrating the metallic upper layers of the SP waveguide (Fig. 2c)[3,29]. Two important factors were considered: First, to ensure a low $|\Delta n|$ and hence weak resonances, the milled slits were made short ($< 1$ $\mu$m) and did not project into the active region. Second, $f_B$ was adjusted by varying $\Lambda$ to iteratively establish the correct relative positions of region (i) of $\rho_{target}$ and the QCL gain. A handful of devices were milled and Fig. 2c shows a typical LCGH-QCL ridge. Figures 2d and e display cross-sectional fundamental mode intensity profiles for SP waveguides with and without upper metal layers, along with their respective $n_{eff}$ values, simulated using FIMMWAVE. The magnitude of $\Delta n$ suggested by the simulations is an overestimate; in reality a smaller value is expected due to the extremely sub-wavelength slit dimension along the propagation axis. However, the simulations do reveal the possibility of a complex $\Delta n$, the imaginary component of which will influence the final lasing mode solutions.

Each device was fully characterized both before and after the introduction of the LCGH, allowing the desired mode switching effect to be clearly discerned. Here we present the outcome of the iterative process, device $A$, which had a measured $\Lambda$ of 13.72 $\mu$m and displayed optimal performance. In contrast to the FP results, the LCGH-QCL spectra recorded from device $A$ revealed operation on six discretely tunable frequencies between 2.860 and 3.024 THz. The selected spectra in Fig. 2f show each mode becoming dominant at a particular driving current density. The switching speed between the six emission conditions is dictated purely by the driving electronics.

In correlating the LCGH spectral response with the tunable emission (Figs. 2b and f), the position of $f_B$, and hence the value of $n_{eff}$ needs to be identified. To this end, the global symmetry of the LCGH response function across $f_B$ was exploited, allowing for a coarse matching of the lasing modes to the LCGH resonances. In device $A$ an approximate symmetry was observed around the mode at 2.970 THz, revealing $n_{eff} \sim 3.68$.

In order to investigate whether the LCGH digital frequency selectivity was both scalable and reproducible, a second FP-QCL, device $B$, was identified with similar FP performance characteristics to device $A$, albeit with a spectral offset of ~80 GHz to lower frequencies. For device $B$, we chose $\Lambda = 14.10$ μm to replicate the relative positions of $f_B$ and QCL gain as in device $A$. This produced tunable laser emission on six dominant frequencies spanning 2.803 to 2.940 THz. Figure 3a displays the LCGH spectra for devices $A$ (blue) and $B$ (red) on a normalized frequency scale. Their mode positions correspond with excellent agreement. Subtle differences do appear between devices, as within each LCGH resonance mode selection occurs in much the same way as in a uniform DFB laser: the round-trip phase defines the exact mode positions[23]. Facet phase was not controlled during FIB milling, nor was the precise value of the complex $\Delta n$, which subtly modified the precise phase solutions. However, the average normalised tuning step in both devices ($\text{TS}_A = 0.011$, $\text{TS}_B = 0.0095$) closely match the target resonance separation ($\text{TS}_{target} = 0.01$), confirming the digital frequency control. Figure 3b displays the driving condition ranges for each LCGH-QCL emission frequency, with devices displaying similar discrete tuning characteristics. Figure 3c shows the measured electrical and optical ($V$-$J$ and $L$-$J$) characteristics of device $B$ before and after LCGH milling. The threshold $J_{th}$ is slightly elevated, but the device retains approximately 70% of the original (pre-FIB) maximum THz power.

Figure 4 shows the LCGH-QCL characteristics of device *C*, which had a similar multi-moded FP performance and the same $\Lambda$ as device *B*, but a deeper milling depth penetrating the top few hundred nanometres of active region, in order to increase $\Delta n$ and the LCGH resonance strengths. Subsequently, it operated on discretely tunable modes at frequencies of 2.815, 2.881 and 2.927 THz. As might be expected, the deeper milling improved the purity of the single-mode emission, with side-mode suppression ratios $\geq 20$ dB after correcting for atmospheric water absorption (Fig. 4a). However, this improved spectral purity came at the expense of a reduced tuning range and a 60% drop in THz power (Fig. 4b). Finally, Fig. 4c shows that each of the three modes dominates over a wide driving condition range.

Beyond discrete tunability, the digital frequency selectivity provided by LCGHs can be harnessed to generate and control arbitrary distributions of lasing modes in THz, as well as shorter wavelength, lasers. Furthermore, although not straightforward given the greater degrees-of-freedom involved, application of LCGH DOFs to the inverse optimisation of photonic and electronic band-structures[30], and the like that utilise properties of wave-like phenomena, may give rise to a host of new materials and devices.

**Acknowledgements**

This work was supported by EPSRC First Grant EP/G064504/1 and partly supported by HMGCC. The authors gratefully acknowledge Dr Ali Gholinia for technical assistance during FIB processing and Dr Michael C. Parker for comments during the manuscript preparation.

**Author Contributions**

S.C. conceived and supervised the project, designed the LCGH, contributed to FIB processing of QCLs, interpreted the data, developed the theoretical framework and prepared the manuscript. O.P.M. fabricated the QCLs, built, characterized and operated the experimental set-up, contributed to FIB processing, discussed the results and implications and contributed to the manuscript at various stages. M.K. contributed to the experimental set-up, carried out measurements and contributed to manuscript preparation. H.E.B. and D.A.R. carried out growth of the QCL.

**Author Information**

The authors declare no competing financial interests. Correspondence and requests for materials should be addressed to S.C. (s.chakraborty@manchester.ac.uk).

**Figure 1 | Principle of a discretely tunable LCGH laser. a**, Symbolic representation of the LCGH. Vertical lines correspond to slits, with a minimum separation Λ. Each horizontal dash represents an additional Λ/2 length. **b**, Calculated LCGH spectral power reflectivity. Shaded area shows the target amplitude envelope covering the first 8 Bragg resonances on either side of the central resonance at $k_B$. Although we could dump the bulk of undesired reflectivity power into only a few Bragg resonances well away from the shaded region (i), we have chosen to position the throw-away region (ii), containing an extra 17 resonances, directly next to region (i). This provided an iterative test and control method for the correct placement of $k_B$ with respect to the QCL gain peak. Inset: Power reflectivity response of a uniform grating. **c**, Detail of 1b. **d**, Illustrative current-controllable QCL gain movement. Note the detuned $k_B$ with respect to the gain peaks. **e**, Illustrative digitally selected lasing modes.

**Figure 2 | Current dependence of the QCL spectra: before and after LCGH introduction. a**, Measured FP laser spectra from device $A$, a 5.72 mm-long THz QCL, at four operating current densities with 10 kHz pulses of 1% duty cycle at 4 K. The 180 μm-wide QCL ridge was defined by wet etching; the upper n[+]-doped GaAs layer (100 nm-thick) was then topped with a shallow palladium-germanium ohmic contact (100 nm) and a titanium/gold overlayer (200 nm). Device characterisation was performed in a three-terminal electrical configuration, using a Janis ST-100 continuous-flow liquid helium cryostat, a nitrogen-purged Bruker Vertex 80 Fourier transform infrared spectrometer (2.2 GHz resolution) and a QMC bolometric detector. Inset: FP THz QCL schematic. **b**, Calculated LCGH spectral power reflectivity (Λ = 13.72 μm, $n_{eff}$ = 3.68, $|\Delta n|$ = 0.1). The shaded region covers the span of the LCGH lasing modes in Figure 2f. **c**, Composite SEM image of the QCL ridge with a 2.77 mm-long, FIB-milled LCGH placed approximately equidistant from each laser facet, written in five stitched sections with a sub-wavelength compound stitching

error. The choice of 100 µm slit width (i.e. narrower than the QCL ridge) enabled simultaneous powering of all ridge sections; the ridge was only wire-bonded at either end (not shown). Simulated cross-sectional fundamental mode intensity profiles within: **d**, an unperturbed SP THz QCL ridge; **e**, an SP THz QCL with the upper metal layers removed. Blue through to yellow indicates low to high intensities. **f**, Pulsed LCGH-QCL spectra (10 kHz, 5% duty cycle) from LCGH device *A* at 4 K. Inset: LCGH-QCL schematic.

**Figure 3 │ Scalability and reproducibility of LCGH-QCL digital frequency selection. a**, Measured LCGH-QCL spectra from devices *A* ($\Lambda$ = 13.72 µm, $f_B$ = 2.973 THz) and *B* (FP cavity length of 6.00 mm, $\Lambda$ = 14.10 µm, $f_B$ = 2.892 THz) on a normalised frequency scale ($n_{eff}$ = 3.68). Within the measurement accuracy of $\Lambda$, the calculated $f_B$ = 2.892 THz for device *B* falls midway between the lasing modes at 2.868 and 2.915 THz, further supporting the correlation between the LCGH response function and the observed spectra. **b**, Driving condition ranges when the six dominant modes in LCGH devices *A* and *B* contain significant power. **c**, *V-J* and *L-J* of FP (black) and LCGH (red) device *B*, collected in pulsed operation (10 kHz, 1% duty cycle) at 4K. Terahertz powers were calibrated to a Thomas Keating absolute THz power meter.

**Figure 4 │ Effect of milling depth on current dependent spectra. a**, Pulsed single-mode laser spectra from LCGH device *C* (FP cavity length of 6.01 mm, $\Lambda$ = 14.10 µm), with deeper milled slits than devices *A* and *B*. **b**, *V-J* and *L-J* of FP (black) and LCGH (red) device *C*, collected in pulsed operation (10 kHz, 1% duty cycle) at 4 K. **c**, Driving condition ranges when the dominant modes in LCGH device *C* contain significant power.

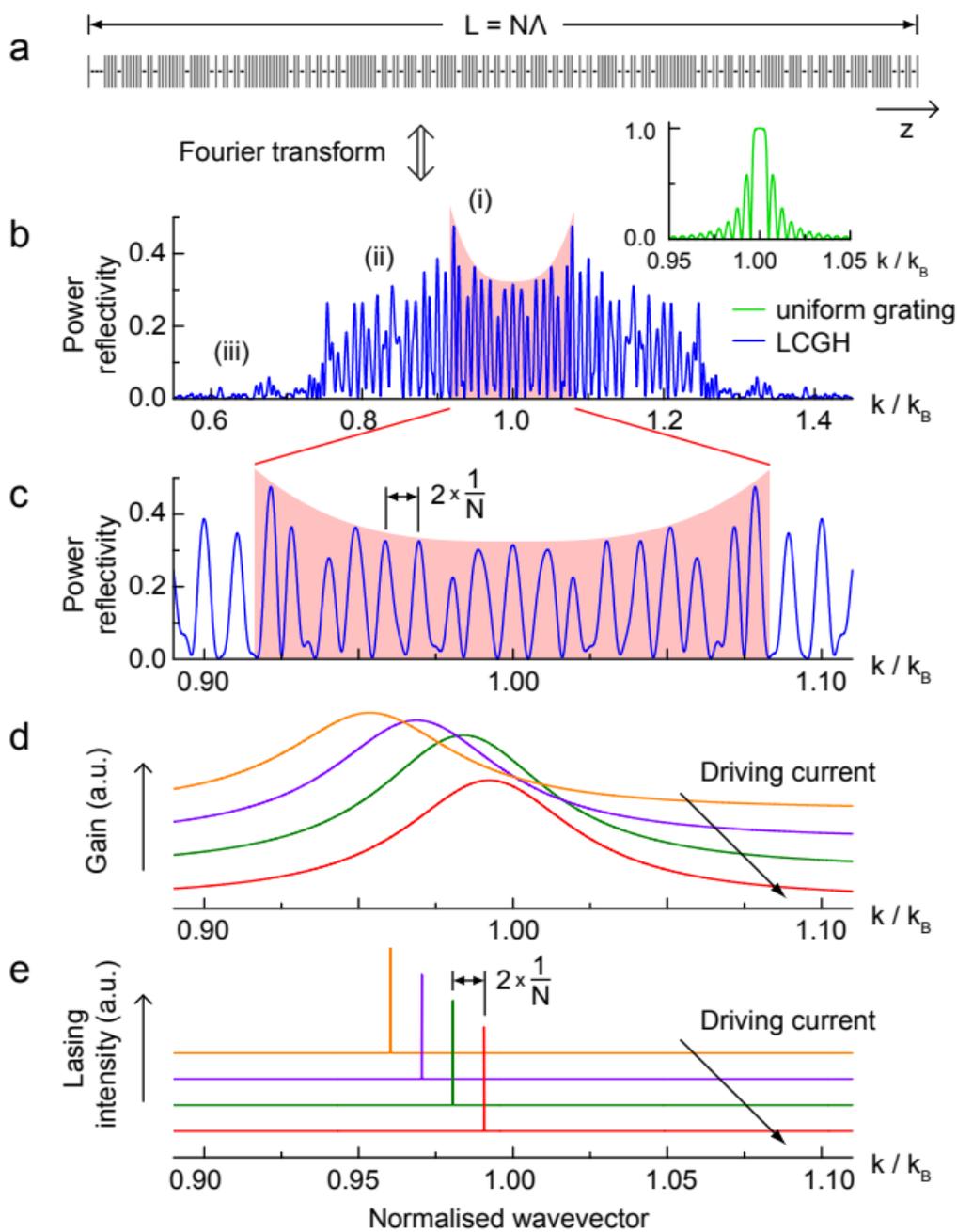

**a** L = NΛ

z

Fourier transform ⇅

**b** Power reflectivity

(i)

(ii)

(iii)

uniform grating
LCGH

k / k_B

1.0

0.0
0.95    1.00    1.05    k / k_B

**c** Power reflectivity

$2 \times \frac{1}{N}$

k / k_B

**d** Gain (a.u.)

Driving current

k / k_B

**e** Lasing intensity (a.u.)

$2 \times \frac{1}{N}$

Driving current

k / k_B

Normalised wavevector

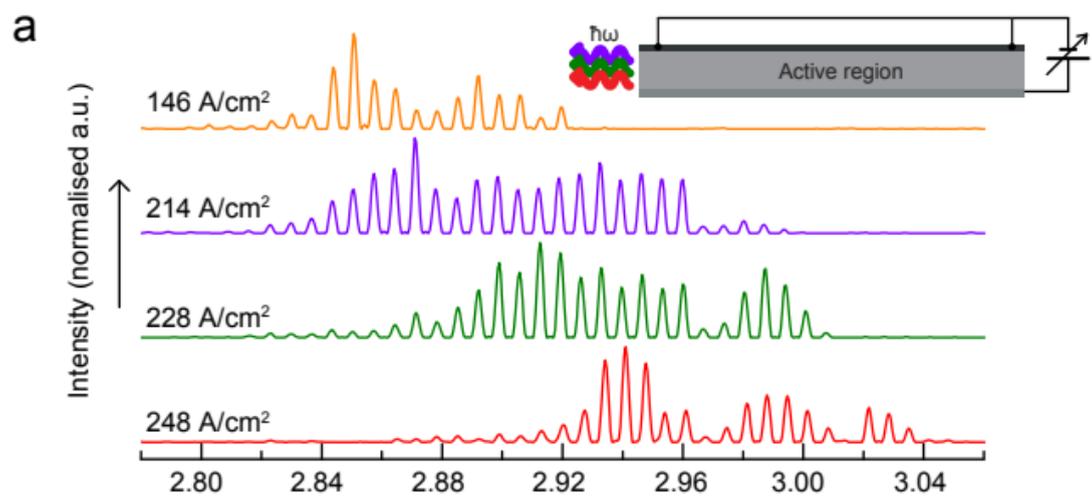

**a**

146 A/cm²

214 A/cm²

228 A/cm²

248 A/cm²

Intensity (normalised a.u.)

**b**

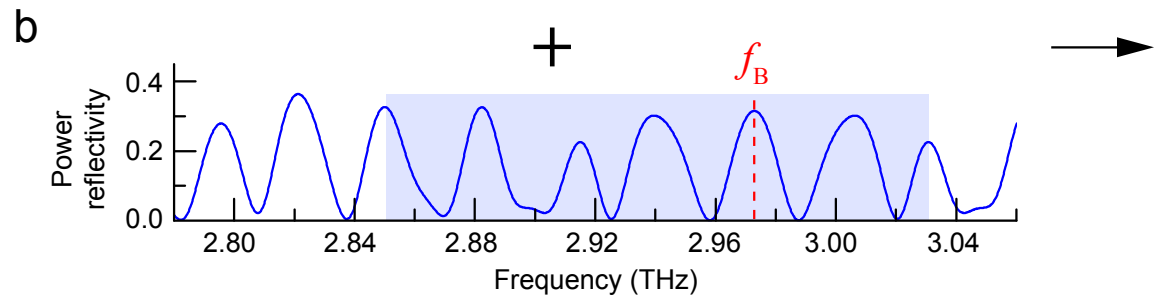

Power reflectivity

$f_B$

Frequency (THz)

**+**

→

**f**

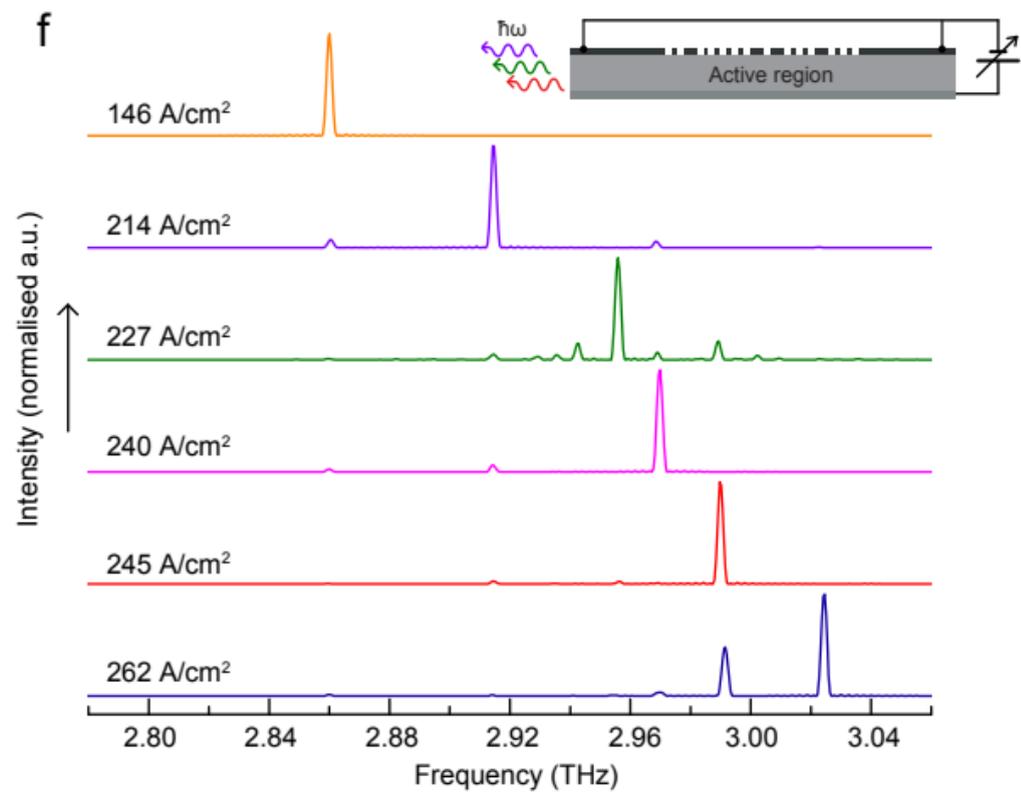

146 A/cm²

214 A/cm²

227 A/cm²

240 A/cm²

245 A/cm²

262 A/cm²

Intensity (normalised a.u.)

Frequency (THz)

**c**

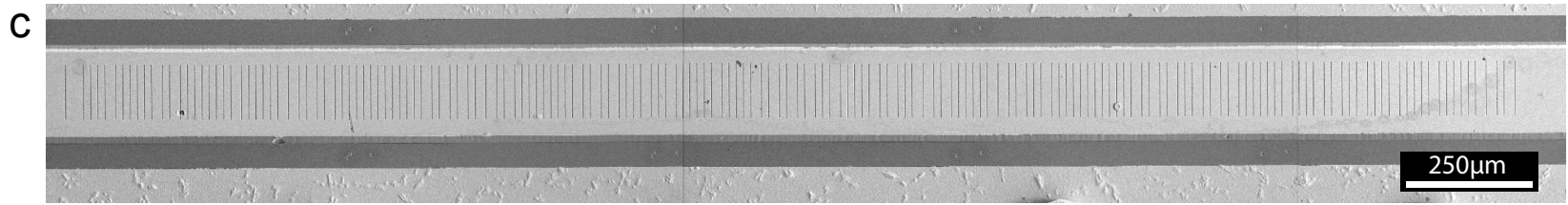

250µm

**d**

$n_{eff} = 3.647 + i0.007$

100µm

**e**

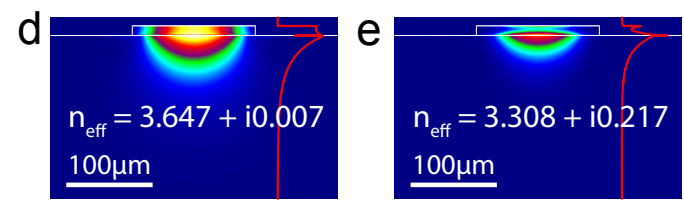

$n_{eff} = 3.308 + i0.217$

100µm

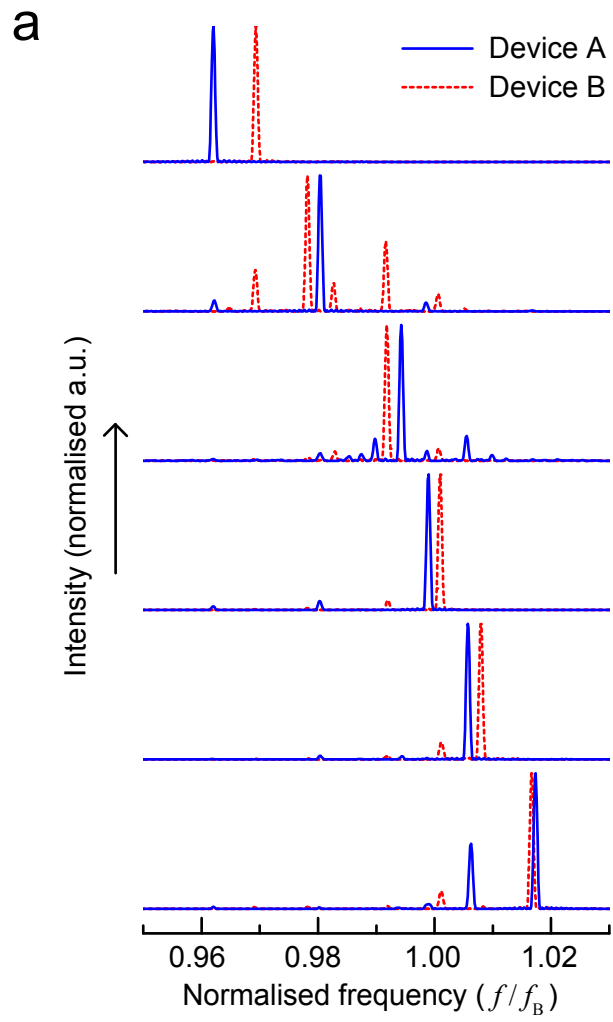

**a**

Intensity (normalised a.u.)

Device A
Device B

Normalised frequency ($f/f_B$)

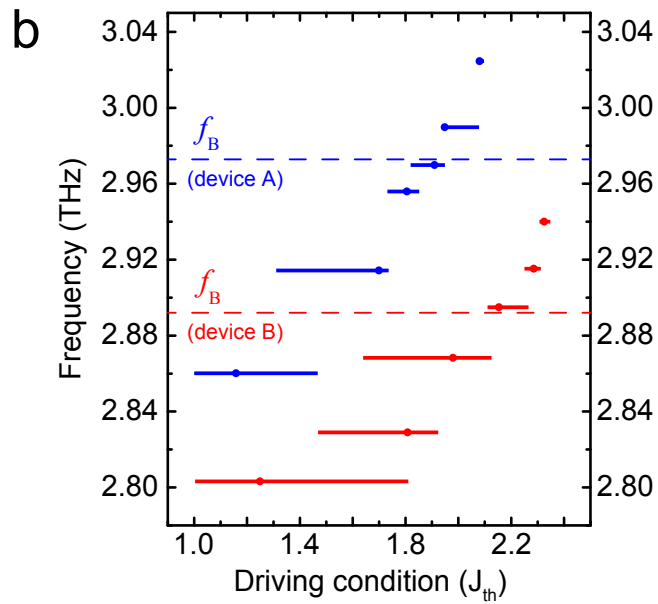

**b**

$f_B$ (device A)

$f_B$ (device B)

Frequency (THz)

Driving condition ($J_B$)

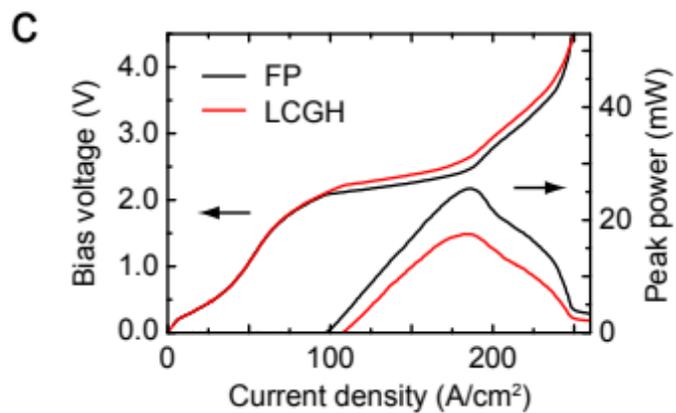

**c**

FP
LCGH

Bias voltage (V)

Peak power (mW)

Current density (A/cm²)

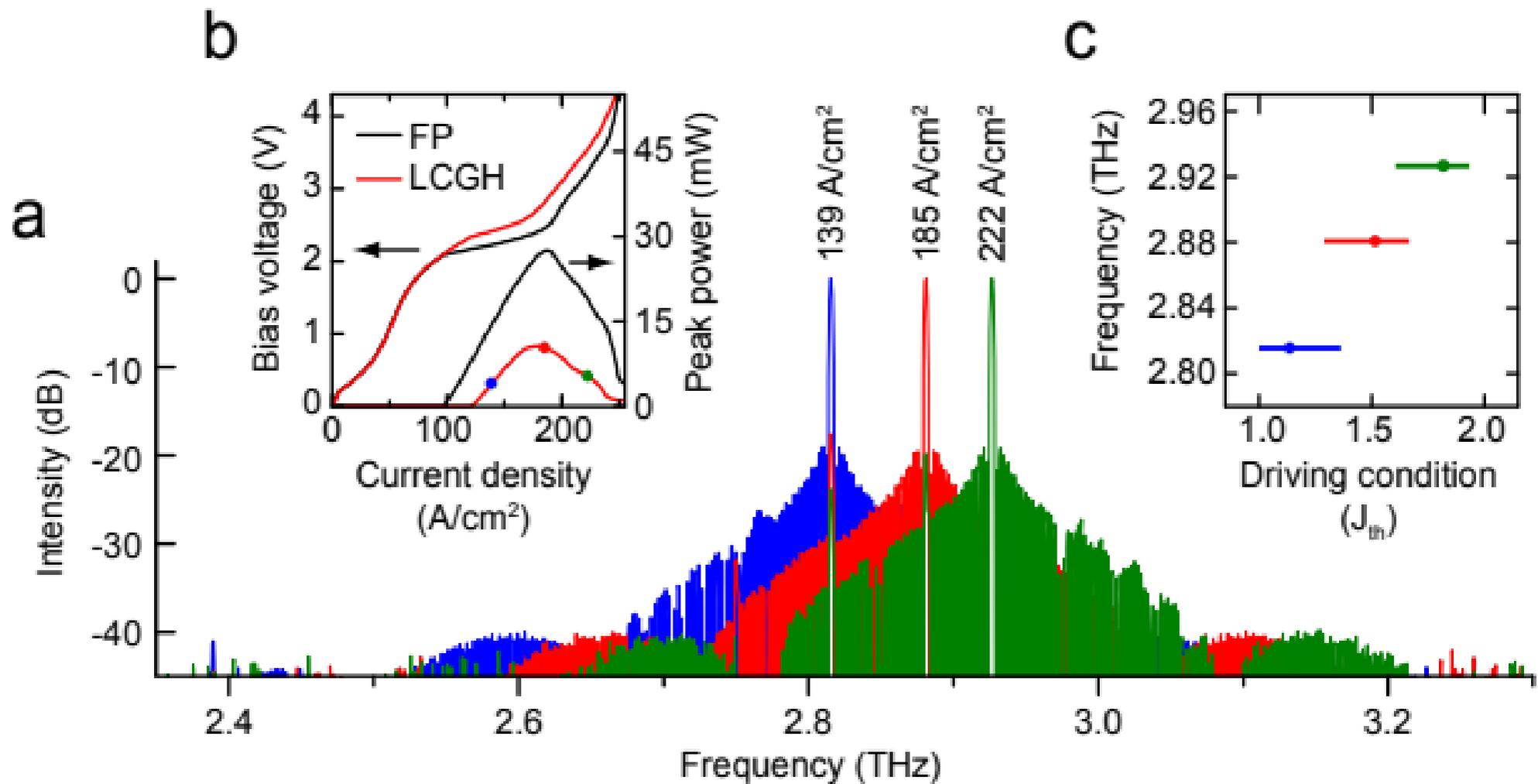

**a**

**b**

**c**

Intensity (dB)

Bias voltage (V) — FP, LCGH

Peak power (mW)

Current density (A/cm²)

Frequency (THz)

Driving condition (J_th)

139 A/cm²   185 A/cm²   222 A/cm²